\documentclass[12pt]{iopart}

\def\PsfigVersion{1.9}
\ifx\undefined\psfig\else\endinput\fi

\let\LaTeXAtSign=\@
\let\@=\relax
\edef\psfigRestoreAt{\catcode`\@=\number\catcode`@\relax}
\catcode`\@=11\relax
\newwrite\@unused
\def\ps@typeout#1{{\let\protect\string\immediate\write\@unused{#1}}}
\ps@typeout{psfig/tex \PsfigVersion}

\def\figurepath{./}

\def\@nnil{\@nil}
\def\@empty{}
\def\@psdonoop#1\@@#2#3{}
\def\@psdo#1:=#2\do#3{\edef\@psdotmp{#2}\ifx\@psdotmp\@empty \else
    \expandafter\@psdoloop#2,\@nil,\@nil\@@#1{#3}\fi}
\def\@psdoloop#1,#2,#3\@@#4#5{\def#4{#1}\ifx #4\@nnil \else
       #5\def#4{#2}\ifx #4\@nnil \else#5\@ipsdoloop #3\@@#4{#5}\fi\fi}
\def\@ipsdoloop#1,#2\@@#3#4{\def#3{#1}\ifx #3\@nnil 
       \let\@nextwhile=\@psdonoop \else
      #4\relax\let\@nextwhile=\@ipsdoloop\fi\@nextwhile#2\@@#3{#4}}
\def\@tpsdo#1:=#2\do#3{\xdef\@psdotmp{#2}\ifx\@psdotmp\@empty \else
    \@tpsdoloop#2\@nil\@nil\@@#1{#3}\fi}
\def\@tpsdoloop#1#2\@@#3#4{\def#3{#1}\ifx #3\@nnil 
       \let\@nextwhile=\@psdonoop \else
      #4\relax\let\@nextwhile=\@tpsdoloop\fi\@nextwhile#2\@@#3{#4}}
\ifx\undefined\fbox
\newdimen\fboxrule
\newdimen\fboxsep
\newdimen\ps@tempdima
\newbox\ps@tempboxa
\long\def\fbox#1{\leavevmode\setbox\ps@tempboxa\hbox{#1}\ps@tempdima\fboxrule
    \advance\ps@tempdima \fboxsep \advance\ps@tempdima \dp\ps@tempboxa
   \hbox{\lower \ps@tempdima\hbox
  {\vbox{\hrule height \fboxrule
          \hbox{\vrule width \fboxrule \hskip\fboxsep
          \vbox{\vskip\fboxsep \box\ps@tempboxa\vskip\fboxsep}\hskip 
                 \fboxsep\vrule width \fboxrule}
                 \hrule height \fboxrule}}}}
\fi
\newread\ps@stream
\newif\ifnot@eof       
\newif\if@noisy        
\newif\if@atend        
\newif\if@psfile       
{\catcode`\%=12\global\gdef\epsf@start{
\def\epsf@PS{PS}
\def\epsf@getbb#1{%
\openin\ps@stream=#1
\ifeof\ps@stream\ps@typeout{Error, File #1 not found}\else
   {\not@eoftrue \chardef\other=12
    \def\do##1{\catcode`##1=\other}\dospecials \catcode`\ =10
    \loop
       \if@psfile
	  \read\ps@stream to \epsf@fileline
       \else{
	  \obeyspaces
          \read\ps@stream to \epsf@tmp\global\let\epsf@fileline\epsf@tmp}
       \fi
       \ifeof\ps@stream\not@eoffalse\else
       \if@psfile\else
       \expandafter\epsf@test\epsf@fileline:. \\%
       \fi
          \expandafter\epsf@aux\epsf@fileline:. \\%
       \fi
   \ifnot@eof\repeat
   }\closein\ps@stream\fi}%
\long\def\epsf@test#1#2#3:#4\\{\def\epsf@testit{#1#2}
			\ifx\epsf@testit\epsf@start\else
\ps@typeout{Warning! File does not start with `\epsf@start'.  It may not be a PostScript file.}
			\fi
			\@psfiletrue} 
{\catcode`\%=12\global\let\epsf@percent=
\long\def\epsf@aux#1#2:#3\\{\ifx#1\epsf@percent
   \def\epsf@testit{#2}\ifx\epsf@testit\epsf@bblit
	\@atendfalse
        \epsf@atend #3 . \\%
	\if@atend	
	   \if@verbose{
		\ps@typeout{psfig: found `(atend)'; continuing search}
	   }\fi
        \else
        \epsf@grab #3 . . . \\%
        \not@eoffalse
        \global\no@bbfalse
        \fi
   \fi\fi}%
\def\epsf@grab #1 #2 #3 #4 #5\\{%
   \global\def\epsf@llx{#1}\ifx\epsf@llx\empty
      \epsf@grab #2 #3 #4 #5 .\\\else
   \global\def\epsf@lly{#2}%
   \global\def\epsf@urx{#3}\global\def\epsf@ury{#4}\fi}%
\def\epsf@atendlit{(atend)} 
\def\epsf@atend #1 #2 #3\\{%
   \def\epsf@tmp{#1}\ifx\epsf@tmp\empty
      \epsf@atend #2 #3 .\\\else
   \ifx\epsf@tmp\epsf@atendlit\@atendtrue\fi\fi}

\chardef\psletter = 11 
\chardef\other = 12

\newif \ifdebug 
\newif\ifc@mpute 
\c@mputetrue 

\let\then = \relax
\def\r@dian{pt }
\let\r@dians = \r@dian
\let\dimensionless@nit = \r@dian
\let\dimensionless@nits = \dimensionless@nit
\def\internal@nit{sp }
\let\internal@nits = \internal@nit
\newif\ifstillc@nverging
\def \Mess@ge #1{\ifdebug \then \message {#1} \fi}

{ 
	\catcode `\@ = \psletter
	\gdef \nodimen {\expandafter \n@dimen \the \dimen}
	\gdef \term #1 #2 #3%
	       {\edef \t@ {\the #1}
		\edef \t@@ {\expandafter \n@dimen \the #2\r@dian}%
		\t@rm {\t@} {\t@@} {#3}%
	       }
	\gdef \t@rm #1 #2 #3%
	       {{%
		\count 0 = 0
		\dimen 0 = 1 \dimensionless@nit
		\dimen 2 = #2\relax
		\Mess@ge {Calculating term #1 of \nodimen 2}%
		\loop
		\ifnum	\count 0 < #1
		\then	\advance \count 0 by 1
			\Mess@ge {Iteration \the \count 0 \space}%
			\Multiply \dimen 0 by {\dimen 2}%
			\Mess@ge {After multiplication, term = \nodimen 0}%
			\Divide \dimen 0 by {\count 0}%
			\Mess@ge {After division, term = \nodimen 0}%
		\repeat
		\Mess@ge {Final value for term #1 of 
				\nodimen 2 \space is \nodimen 0}%
		\xdef \Term {#3 = \nodimen 0 \r@dians}%
		\aftergroup \Term
	       }}
	\catcode `\p = \other
	\catcode `\t = \other
	\gdef \n@dimen #1pt{#1} 
}

\def \Divide #1by #2{\divide #1 by #2} 

\def \Multiply #1by #2
       {{
	\count 0 = #1\relax
	\count 2 = #2\relax
	\count 4 = 65536
	\Mess@ge {Before scaling, count 0 = \the \count 0 \space and
			count 2 = \the \count 2}%
	\ifnum	\count 0 > 32767 
	\then	\divide \count 0 by 4
		\divide \count 4 by 4
	\else	\ifnum	\count 0 < -32767
		\then	\divide \count 0 by 4
			\divide \count 4 by 4
		\else
		\fi
	\fi
	\ifnum	\count 2 > 32767 
	\then	\divide \count 2 by 4
		\divide \count 4 by 4
	\else	\ifnum	\count 2 < -32767
		\then	\divide \count 2 by 4
			\divide \count 4 by 4
		\else
		\fi
	\fi
	\multiply \count 0 by \count 2
	\divide \count 0 by \count 4
	\xdef \product {#1 = \the \count 0 \internal@nits}%
	\aftergroup \product
       }}

\def\r@duce{\ifdim\dimen0 > 90\r@dian \then   
		\multiply\dimen0 by -1
		\advance\dimen0 by 180\r@dian
		\r@duce
	    \else \ifdim\dimen0 < -90\r@dian \then  
		\advance\dimen0 by 360\r@dian
		\r@duce
		\fi
	    \fi}

\def\Sine#1%
       {{%
	\dimen 0 = #1 \r@dian
	\r@duce
	\ifdim\dimen0 = -90\r@dian \then
	   \dimen4 = -1\r@dian
	   \c@mputefalse
	\fi
	\ifdim\dimen0 = 90\r@dian \then
	   \dimen4 = 1\r@dian
	   \c@mputefalse
	\fi
	\ifdim\dimen0 = 0\r@dian \then
	   \dimen4 = 0\r@dian
	   \c@mputefalse
	\fi
	\ifc@mpute \then
		\divide\dimen0 by 180
		\dimen0=3.141592654\dimen0
		\dimen 2 = 3.1415926535897963\r@dian 
		\divide\dimen 2 by 2 
		\Mess@ge {Sin: calculating Sin of \nodimen 0}%
		\count 0 = 1 
		\dimen 2 = 1 \r@dian 
		\dimen 4 = 0 \r@dian 
		\loop
			\ifnum	\dimen 2 = 0 
			\then	\stillc@nvergingfalse 
			\else	\stillc@nvergingtrue
			\fi
			\ifstillc@nverging 
			\then	\term {\count 0} {\dimen 0} {\dimen 2}%
				\advance \count 0 by 2
				\count 2 = \count 0
				\divide \count 2 by 2
				\ifodd	\count 2 
				\then	\advance \dimen 4 by \dimen 2
				\else	\advance \dimen 4 by -\dimen 2
				\fi
		\repeat
	\fi		
			\xdef \sine {\nodimen 4}%
       }}

\def\Cosine#1{\ifx\sine\UnDefined\edef\Savesine{\relax}\else
		             \edef\Savesine{\sine}\fi
	{\dimen0=#1\r@dian\advance\dimen0 by 90\r@dian
	 \Sine{\nodimen 0}
	 \xdef\cosine{\sine}
	 \xdef\sine{\Savesine}}}	      

\def\psdraft{
	\def\@psdraft{0}
}
\def\psfull{
	\def\@psdraft{100}
}

\psfull

\newif\if@scalefirst
\def\psscalefirst{\@scalefirsttrue}
\def\psrotatefirst{\@scalefirstfalse}
\psrotatefirst

\newif\if@draftbox
\def\psnodraftbox{
	\@draftboxfalse
}
\def\psdraftbox{
	\@draftboxtrue
}
\@draftboxtrue

\newif\if@prologfile
\newif\if@postlogfile
\def\pssilent{
	\@noisyfalse
}
\def\psnoisy{
	\@noisytrue
}
\psnoisy
\newif\if@bbllx
\newif\if@bblly
\newif\if@bburx
\newif\if@bbury
\newif\if@height
\newif\if@width
\newif\if@rheight
\newif\if@rwidth
\newif\if@angle
\newif\if@clip
\newif\if@verbose
\def\@p@@sclip#1{\@cliptrue}

\newif\if@decmpr

\def\@p@@sfigure#1{\def\@p@sfile{null}\def\@p@sbbfile{null}
	        \openin1=#1.bb
		\ifeof1\closein1
	        	\openin1=\figurepath#1.bb
			\ifeof1\closein1
			        \openin1=#1
				\ifeof1\closein1%
				       \openin1=\figurepath#1
					\ifeof1
					   \ps@typeout{Error, File #1 not found}
						\if@bbllx\if@bblly
				   		\if@bburx\if@bbury
			      				\def\@p@sfile{#1}%
			      				\def\@p@sbbfile{#1}%
							\@decmprfalse
				  	   	\fi\fi\fi\fi
					\else\closein1
				    		\def\@p@sfile{\figurepath#1}%
				    		\def\@p@sbbfile{\figurepath#1}%
						\@decmprfalse
	                       		\fi%
			 	\else\closein1%
					\def\@p@sfile{#1}
					\def\@p@sbbfile{#1}
					\@decmprfalse
			 	\fi
			\else
				\def\@p@sfile{\figurepath#1}
				\def\@p@sbbfile{\figurepath#1.bb}
				\@decmprtrue
			\fi
		\else
			\def\@p@sfile{#1}
			\def\@p@sbbfile{#1.bb}
			\@decmprtrue
		\fi}

\def\@p@@sfile#1{\@p@@sfigure{#1}}

\def\@p@@sbbllx#1{
		\@bbllxtrue
		\dimen100=#1
		\edef\@p@sbbllx{\number\dimen100}
}
\def\@p@@sbblly#1{
		\@bbllytrue
		\dimen100=#1
		\edef\@p@sbblly{\number\dimen100}
}
\def\@p@@sbburx#1{
		\@bburxtrue
		\dimen100=#1
		\edef\@p@sbburx{\number\dimen100}
}
\def\@p@@sbbury#1{
		\@bburytrue
		\dimen100=#1
		\edef\@p@sbbury{\number\dimen100}
}
\def\@p@@sheight#1{
		\@heighttrue
		\dimen100=#1
   		\edef\@p@sheight{\number\dimen100}
}
\def\@p@@swidth#1{
		\@widthtrue
		\dimen100=#1
		\edef\@p@swidth{\number\dimen100}
}
\def\@p@@srheight#1{
		\@rheighttrue
		\dimen100=#1
		\edef\@p@srheight{\number\dimen100}
}
\def\@p@@srwidth#1{
		\@rwidthtrue
		\dimen100=#1
		\edef\@p@srwidth{\number\dimen100}
}
\def\@p@@sangle#1{
		\@angletrue
		\edef\@p@sangle{#1} 
}
\def\@p@@ssilent#1{ 
		\@verbosefalse
}
\def\@p@@sprolog#1{\@prologfiletrue\def\@prologfileval{#1}}
\def\@p@@spostlog#1{\@postlogfiletrue\def\@postlogfileval{#1}}
\def\@cs@name#1{\csname #1\endcsname}
\def\@setparms#1=#2,{\@cs@name{@p@@s#1}{#2}}
\def\ps@init@parms{
		\@bbllxfalse \@bbllyfalse
		\@bburxfalse \@bburyfalse
		\@heightfalse \@widthfalse
		\@rheightfalse \@rwidthfalse
		\def\@p@sbbllx{}\def\@p@sbblly{}
		\def\@p@sbburx{}\def\@p@sbbury{}
		\def\@p@sheight{}\def\@p@swidth{}
		\def\@p@srheight{}\def\@p@srwidth{}
		\def\@p@sangle{0}
		\def\@p@sfile{} \def\@p@sbbfile{}
		\def\@p@scost{10}
		\def\@sc{}
		\@prologfilefalse
		\@postlogfilefalse
		\@clipfalse
		\if@noisy
			\@verbosetrue
		\else
			\@verbosefalse
		\fi
}
\def\parse@ps@parms#1{
	 	\@psdo\@psfiga:=#1\do
		   {\expandafter\@setparms\@psfiga,}}
\newif\ifno@bb
\def\bb@missing{
	\if@verbose{
		\ps@typeout{psfig: searching \@p@sbbfile \space  for bounding box}
	}\fi
	\no@bbtrue
	\epsf@getbb{\@p@sbbfile}
        \ifno@bb \else \bb@cull\epsf@llx\epsf@lly\epsf@urx\epsf@ury\fi
}	
\def\bb@cull#1#2#3#4{
	\dimen100=#1 bp\edef\@p@sbbllx{\number\dimen100}
	\dimen100=#2 bp\edef\@p@sbblly{\number\dimen100}
	\dimen100=#3 bp\edef\@p@sbburx{\number\dimen100}
	\dimen100=#4 bp\edef\@p@sbbury{\number\dimen100}
	\no@bbfalse
}
\newdimen\p@intvaluex
\newdimen\p@intvaluey
\def\rotate@#1#2{{\dimen0=#1 sp\dimen1=#2 sp
		  \global\p@intvaluex=\cosine\dimen0
		  \dimen3=\sine\dimen1
		  \global\advance\p@intvaluex by -\dimen3
		  \global\p@intvaluey=\sine\dimen0
		  \dimen3=\cosine\dimen1
		  \global\advance\p@intvaluey by \dimen3
		  }}
\def\compute@bb{
		\no@bbfalse
		\if@bbllx \else \no@bbtrue \fi
		\if@bblly \else \no@bbtrue \fi
		\if@bburx \else \no@bbtrue \fi
		\if@bbury \else \no@bbtrue \fi
		\ifno@bb \bb@missing \fi
		\ifno@bb \ps@typeout{FATAL ERROR: no bb supplied or found}
			\no-bb-error
		\fi
		\count203=\@p@sbburx
		\count204=\@p@sbbury
		\advance\count203 by -\@p@sbbllx
		\advance\count204 by -\@p@sbblly
		\edef\ps@bbw{\number\count203}
		\edef\ps@bbh{\number\count204}
		\if@angle 
			\Sine{\@p@sangle}\Cosine{\@p@sangle}
	        	{\dimen100=\maxdimen\xdef\r@p@sbbllx{\number\dimen100}
					    \xdef\r@p@sbblly{\number\dimen100}
			                    \xdef\r@p@sbburx{-\number\dimen100}
					    \xdef\r@p@sbbury{-\number\dimen100}}
                        \def\minmaxtest{
			   \ifnum\number\p@intvaluex<\r@p@sbbllx
			      \xdef\r@p@sbbllx{\number\p@intvaluex}\fi
			   \ifnum\number\p@intvaluex>\r@p@sbburx
			      \xdef\r@p@sbburx{\number\p@intvaluex}\fi
			   \ifnum\number\p@intvaluey<\r@p@sbblly
			      \xdef\r@p@sbblly{\number\p@intvaluey}\fi
			   \ifnum\number\p@intvaluey>\r@p@sbbury
			      \xdef\r@p@sbbury{\number\p@intvaluey}\fi
			   }
			\rotate@{\@p@sbbllx}{\@p@sbblly}
			\minmaxtest
			\rotate@{\@p@sbbllx}{\@p@sbbury}
			\minmaxtest
			\rotate@{\@p@sbburx}{\@p@sbblly}
			\minmaxtest
			\rotate@{\@p@sbburx}{\@p@sbbury}
			\minmaxtest
			\edef\@p@sbbllx{\r@p@sbbllx}\edef\@p@sbblly{\r@p@sbblly}
			\edef\@p@sbburx{\r@p@sbburx}\edef\@p@sbbury{\r@p@sbbury}
		\fi
		\count203=\@p@sbburx
		\count204=\@p@sbbury
		\advance\count203 by -\@p@sbbllx
		\advance\count204 by -\@p@sbblly
		\edef\@bbw{\number\count203}
		\edef\@bbh{\number\count204}
}
\def\in@hundreds#1#2#3{\count240=#2 \count241=#3
		     \count100=\count240	
		     \divide\count100 by \count241
		     \count101=\count100
		     \multiply\count101 by \count241
		     \advance\count240 by -\count101
		     \multiply\count240 by 10
		     \count101=\count240	
		     \divide\count101 by \count241
		     \count102=\count101
		     \multiply\count102 by \count241
		     \advance\count240 by -\count102
		     \multiply\count240 by 10
		     \count102=\count240	
		     \divide\count102 by \count241
		     \count200=#1\count205=0
		     \count201=\count200
			\multiply\count201 by \count100
		 	\advance\count205 by \count201
		     \count201=\count200
			\divide\count201 by 10
			\multiply\count201 by \count101
			\advance\count205 by \count201
		     \count201=\count200
			\divide\count201 by 100
			\multiply\count201 by \count102
			\advance\count205 by \count201
		     \edef\@result{\number\count205}
}
\def\compute@wfromh{
		\in@hundreds{\@p@sheight}{\@bbw}{\@bbh}
		\edef\@p@swidth{\@result}
}
\def\compute@hfromw{
	        \in@hundreds{\@p@swidth}{\@bbh}{\@bbw}
		\edef\@p@sheight{\@result}
}
\def\compute@handw{
		\if@height 
			\if@width
			\else
				\compute@wfromh
			\fi
		\else 
			\if@width
				\compute@hfromw
			\else
				\edef\@p@sheight{\@bbh}
				\edef\@p@swidth{\@bbw}
			\fi
		\fi
}
\def\compute@resv{
		\if@rheight \else \edef\@p@srheight{\@p@sheight} \fi
		\if@rwidth \else \edef\@p@srwidth{\@p@swidth} \fi
}
\def\compute@sizes{
	\compute@bb
	\if@scalefirst\if@angle
	\if@width
	   \in@hundreds{\@p@swidth}{\@bbw}{\ps@bbw}
	   \edef\@p@swidth{\@result}
	\fi
	\if@height
	   \in@hundreds{\@p@sheight}{\@bbh}{\ps@bbh}
	   \edef\@p@sheight{\@result}
	\fi
	\fi\fi
	\compute@handw
	\compute@resv}

\def\psfig#1{\vbox {
	\ps@init@parms
	\parse@ps@parms{#1}
	\compute@sizes
	\ifnum\@p@scost<\@psdraft{
		\special{ps::[begin] 	\@p@swidth \space \@p@sheight \space
				\@p@sbbllx \space \@p@sbblly \space
				\@p@sbburx \space \@p@sbbury \space
				startTexFig \space }
		\if@angle
			\special {ps:: \@p@sangle \space rotate \space} 
		\fi
		\if@clip{
			\if@verbose{
				\ps@typeout{(clip)}
			}\fi
			\special{ps:: doclip \space }
		}\fi
		\if@prologfile
		    \special{ps: plotfile \@prologfileval \space } \fi
		\if@decmpr{
			\if@verbose{
				\ps@typeout{psfig: including \@p@sfile.Z \space }
			}\fi
			\special{ps: plotfile "`zcat \@p@sfile.Z" \space }
		}\else{
			\if@verbose{
				\ps@typeout{psfig: including \@p@sfile \space }
			}\fi
			\special{ps: plotfile \@p@sfile \space }
		}\fi
		\if@postlogfile
		    \special{ps: plotfile \@postlogfileval \space } \fi
		\special{ps::[end] endTexFig \space }
		\vbox to \@p@srheight sp{
			\hbox to \@p@srwidth sp{
				\hss
			}
		\vss
		}
	}\else{
		\if@draftbox{		
			\hbox{\frame{\vbox to \@p@srheight sp{
			\vss
			\hbox to \@p@srwidth sp{ \hss \@p@sfile \hss }
			\vss
			}}}
		}\else{
			\vbox to \@p@srheight sp{
			\vss
			\hbox to \@p@srwidth sp{\hss}
			\vss
			}
		}\fi

	}\fi
}}
\psfigRestoreAt
\let\@=\LaTeXAtSign

\begin{document}

\title{Can gamma-ray bursts constrain quintessence?}

\author{Tristano Di Girolamo\dag\footnote[3]{To whom correspondence should 
be addressed (tristano@na.infn.it)} 
Riccardo Catena\ddag\, Mario Vietri\ddag\, and Giuseppe Di Sciascio\dag}

\address{\dag\ Istituto Nazionale di Fisica Nucleare, Sezione di Napoli, 
Complesso Universitario di Monte Sant'Angelo, Via Cintia, 80126 Napoli, Italy}

\address{\ddag\ Scuola Normale Superiore, Piazza dei Cavalieri, 56100 Pisa, 
Italy}

\begin{abstract}
Using the narrow clustering of the geometrically corrected
gamma-ray energies released by gamma-ray bursts, we investigate
the possibility of using these sources as standard candles to
probe cosmological parameters such as the matter density $\Omega_m$ 
and the cosmological constant energy density $\Omega_{\Lambda}$.
By simulating different samples of gamma-ray bursts, we find that
$\Omega_m $ can be determined with accuracy
$\sim$7\% with data from 300 sources. We also show that, if $\Omega=1$ 
is due to a quintessence field, some of the models
proposed in the literature may be discriminated from a Universe
with cosmological constant, by a similar--sized sample of gamma-ray bursts.

\end{abstract}




\section{Introduction}
Recent studies have pointed out that Gamma-Ray Bursts (GRBs) may
be used as standard cosmological candles. The prompt $\gamma$-ray
energy release, when neglect is made of the conical geometry of
the emission, spans nearly three orders of magnitude, and the
distribution of the opening angles of the emission, as deduced
from the timing of the achromatic steepening of the afterglow
emission, spans an identically wide range of values. However, when
the apparently isotropic energy release and the conic opening of
the emission are combined to infer the intrinsic, true energy
release, the resulting distribution does not widen, as is expected
for uncorrelated data, but shrinks to a very well determined value
(Frail et al. 2001; Panaitescu \& Kumar 2001; Bloom, Frail, \& Kulkarni 2003), 
with a remarkably small (one--sided) scattering, corresponding to about a 
factor of $2$ in total energy. Similar studies in the X--ray band (Piran et al.
2001; Berger, Kulkarni, \& Frail 2003) have reproduced the same
results.

It is thus very tempting to study to what extent this property of
GRBs makes them suitable cosmological standard candles. After an early
investigation made by Cohen \& Piran (1997), Schaefer
(2003) proposed using two well known correlations of the
GRBs luminosity (with variability, and with time delay) to the
same end, while Dai, Liang, \& Xu (2004) and Ghirlanda et al. (2004)
exploited the recently reported relationship between the beaming--corrected 
$\gamma$-ray energy and the local observer peak energy in GRBs 
(Ghirlanda, Ghisellini, \& Lazzati 2004). 
We instead neglect these three relationships and
concentrate on the very narrow spread of the true,
geometrically corrected energy release as a distance indicator,
recalling however that its determination for 
any given bursts requires substantially more information than the methods 
presented by Schaefer, and the other authors mentioned above.

As for the possible variation of ambient density from
burst to burst, which may widen the distribution of bursts energies,
Frail et al. (2001) remarked that this spread is already
contained in their data sample, and yet the distribution of energy
releases is still very narrow. If we were somehow able to measure
the distribution ambient densities, and subtract these from the sample,
the distribution of energy releases should narrow even more, not widen: 
in fact, since we obviously expect the two distributions to be uncorrelated, 
we also expect the one resulting from their combination to be wider than 
the intrinsic distribution of energy releases. 

There are at least two respects in which GRBs are better than type
Ia SuperNovae (SNIa) as cosmological candles, one in which they are worse, 
and one in which they are probably even. 
On the one hand, GRBs are easy to find and locate:
even 1980s' technology allowed BATSE to locate $\sim$1 GRB per day,
despite an incompleteness of about $1/3$, making the build--up of
a 300--object database a one--year enterprise, with old technology.
The launch of the {\it Swift} satellite, which took place on 20 November 2004,
is expected to detect GRBs at about the same rate as BATSE, but
with a nearly perfect capacity for identifying their redshifts
simultaneously with the afterglow observations
\footnote{http://swift.gsfc.nasa.gov/docs/swift/proposals/appendix\_f.html}.
Second, GRBs have been detected out to very high redshifts: even the current 
sample of about 40 objects (Greiner 2004)
contains several events with $z> 3$, with an absolute maximum of
$z = 4.5$ for GRB 000131. This should
be contrasted with the difficulty of locating SNe with $z > 1$, 
and the absolute lack of any SN with $z > 2$. The currently
observed distribution of GRBs redshifts contains instead 21
events with $z>1$ out of a total of 39 (see Figure~\ref{plotfour}).

On the other hand, the distribution of luminosities of SNIa
is narrower than the distribution of GRBs energy releases,
corresponding to a magnitude dispersion
$\sigma_M = 0.18$ rather than $\sigma_M = 0.75$.
However, the two break even (probably) in terms of our
understanding of the underlying physical reasons for the
uniformity of the distributions, which is wanting in both cases.

Thus GRBs may provide a complementary standard candle, out to
distances which cannot be probed by SNIa, their major
limitation being the larger intrinsic scatter of the energy
release, as compared to the small scatter in peak luminosities of
SNIa. It is thus important to assess whether this larger
scatter still allows GRBs to be used as standard candles. To this
end, and as a first aim of the paper, we carry out numerical
simulations of random samples of GRBs, whose energy releases are
distributed as found out by Frail et al. (2001), to see to what
extent global cosmological parameters can be identified by an
arbitrarily large (but within reason) sample of hypothetical
observations.

As a second aim of the paper, we study, also by means of
simulations, whether the larger redshift range spanned by GRBs,
when compared with SNIa, allows us to identify specific
models for quintessence. If the non--matter component of the
overall energy density in the Universe were indeed a constant, at
$z\approx 1$ the increase in the matter content would dwarf it,
and there would be no difference, at larger redshifts, between a
model with cosmological constant, and one without (we call this
{\it the null hypothesis}). However, if the cosmological constant
is not constant at all, but is provided by the new heuristic field
called quintessence, one may hope that at least some models
display evolution of the cosmological distances (luminosity,
fluence, angular, and so on), which differ substantially from
those of the model with cosmological constant. It is thus our
second aim to study universes with different, simple models
for quintessence, to see whether GRBs observations may be
able to discriminate between them. In other words, we study
whether GRBs can reject the hypothesis of a {\it constant}
$\Lambda$.

We stress that this paper is not aimed at displaying the potential
for cosmological investigation by any coming satellite, but
instead at determining whether the size of a realistically
obtainable set of data (perhaps to be obtained by means of a
dedicated satellite) is useful for cosmological studies. We assume
that we know, for every burst in our sample, the redshift, and the
opening--corrected apparent fluence ({\it i.e.}, the apparent luminosity
integrated over the burst duration), and that there is no evolution 
with redshift of the bursts intrinsic energy release. We remark that it is 
{\it not} necessary to have a complete and homogeneous sample of
objects to carry out this exercise, and that the precise value of
the bursts average energy release is not necessary, because as
usual in cosmological tests, we are fitting the dependence of the
luminosity distance upon redshift and cosmological parameters, not
its absolute normalization.

The plan of the paper is as follows. In Section 2, we display a
simulation with the simple aim of showing the power of a set of
300 GRBs distributed out to large redshifts, in rejecting or
accepting the presence of a cosmological constant term in the
Universe density distribution. A test like this would also be
useful in practice, since it would be completely independent of
observations of fluctuations in the Cosmic Microwave Background
Radiation (CMBR). Then, in Section 3, we
assume a $\Omega_{\Lambda} = 0$, $\Omega_m = 1$ cosmology, and test the
ability of similar--size sets of GRBs to determine $\Omega_m = 1$.
In Section 4, we assume instead $\Omega_{\Lambda} \neq 0$,
$\Omega_m+\Omega_{\Lambda} = 1$, and test
the ability of the same samples of GRBs to identify the correct
values of $\Omega_m$. In Section 5, we abandon the hypothesis that
$\Lambda$ is a constant, and turn to different quintessence
models, showing that at least one of the important ones
(Gasperini, Piazza, \& Veneziano 2002) can be easily discriminated
from the others, and from the null hypothesis. In Section 6, we
summarize and conclude.

\section{A simple test}
First, in order to show what we are 
aiming at, we performed a Kolmogorov-Smirnov (KS) test on two data sets 
made of 300 GRBs simulated in two different cosmological models, one with
$\Omega_m = 1$ and $\Omega_{\Lambda} = 0$ and the other with 
$\Omega_m = 0.3$ and $\Omega_{\Lambda} = 0.7$, but both with 
a Hubble constant $H_0 = 65$ km s$^{-1}$ Mpc$^{-1}$ (as it will be 
assumed throughout the paper).
We preferred the KS test to others since it is applicable to any kind of
continuous distribution that is a function of a single independent variable,
which is the case we are dealing with. The $\chi^2$ test, for comparison,
is more suited to point out differences between binned distributions.
For the KS test each list of data points, after ordering, is converted to a 
cumulative distribution function giving the fraction of data points to the 
left of a given value for the variable. Then the maximum value $D$ of the 
absolute difference between these two cumulative distribution functions is 
adopted as the test statistic, and the probability $Q_{KS}$ of finding values 
greater than the observed $D$ gives the significance level for the null 
hypothesis that the data sets are drawn from the same distribution
(Von Mises 1964).
 
We assume that GRBs are indeed standard candles with true 
$\gamma$-ray energy released, $E_{\gamma} $, following a Gaussian distribution 
in its logarithm with mean $\mu = 51.1$ (if $E_{\gamma} $ is expressed in 
erg units, Bloom, Frail, \& Kulkarni 2003) and $\sigma =0.3$ (corresponding to 
a multiplicative factor of 2), and that they are distributed in the Universe
according to the model of star formation rate
$R_{SF1} (z)$ reported in Porciani \& Madau (2001), which matches the 
$\log N - \log P$ relation (GRB number counts vs. peak photon
flux) obtained with BATSE data. Applying the KS test to the redshift
distributions, we found that the probability that the two data sets are
the same is $Q_{KS}=0.031$, a ``no man's land'' value for this test. On the
other hand, the application of the KS test to the parameter 
$\log d_L^2 (z)$, where $d_L (z)$ is the luminosity distance, resulted in a
significant probability $Q_{KS} \sim 10^{-14}$, which tells us that it is 
possible to discriminate between the two different cosmological models if a
set of 300 GRB luminosity distances is known, 
without any reference to CMBR data.

\section{Simulations in a $\Lambda =0$ cosmology}
We consider now a $\Lambda =0$ 
cosmology, in which the only contribution to the density parameter is given 
by $\Omega_m$. We assume for GRBs the same energy distribution as for the KS
test. However, the assumed mean value is not relevant for our
investigation, since it is the dispersion value that constrains the
cosmological density parameter. The dispersion of the $\gamma$-ray energy 
released in GRBs may be pinned down in the future by a local sample of 
sources, such as the recently discovered GRB 030329 and 031203, at 
$z=0.1685$ (Price et al. 2003) and $z=0.1055$ 
(Malesani et al. 2004) respectively (see Section 5 for a more detailed 
discussion on this point).

The standard candle energy is related to the fluence of the 
burst $f_{\gamma} = E_{\gamma} (1+z)/(4\pi d_L^2 (z))$
via the luminosity distance $d_L (z)$, whose expression for $\Lambda =0$ is:
\begin{equation}
d_L (z) = \frac{c}{H_0} \frac{2\left[ 2-\Omega_m +\Omega_m z -(2-\Omega_m )
 \sqrt{1+\Omega_m z} \right] }{\Omega_m^2 }
\end{equation}
Since the $k$-correction is independent of any cosmological parameter,
we take no account of it. A discussion about its effects on the
distribution of GRBs energy releases is made in Bloom, Frail, \& Sari (2001). 
In order to have a linear propagation of errors throughout our 
simulations, we choose to construct with GRBs a Hubble diagram 
$\log d_L^2 - z$, since the distribution of the parameter $\log d_L^2 $ is 
the same of that of $\log E_{\gamma}$, and therefore it is Gaussian.  

The number of GRBs per redshift unit is given by the expression:
\begin{equation}
\frac{dN_{GRB}}{dz} = \frac{n(z)(dV/dz)}{\int_0^{z_{max}} n(z)(dV/dz) dz}
\label{nzdistr}
\end{equation}
where $n(z)$ is the redshift distribution function, extending to the
maximum redshift for GRB explosions $z_{max}$, and $dV/dz$ is the
comoving volume element, which for $\Lambda =0$ is:
\begin{equation}
\frac{dV}{dz} = \frac{c}{H_0} \frac{4\pi d_L^2 (z)}
 {(1+z)^2 \left[\Omega_m (1+z)^3 +(1-\Omega_m )(1+z)^2 \right]^{1/2}}
\end{equation} 
As for the KS test, we assume that the redshifts of GRBs are distributed 
according to the model of star formation rate $R_{SF1} (z)$ reported in 
Porciani \& Madau (2001):
\begin{equation}
n(z) \propto \frac{\exp (3.4z)}{\exp (3.8z)+45} 
\label{madaufunc}
\end{equation}
This function increases rapidly between $z=0$ and 1, peaks between $z=1$ and 
2, and gently declines at higher redshifts. We fix $z_{max} =5$.

In order to study the ability of GRBs in probing the cosmological 
parameters as a function of their number, we have simulated different samples 
with $N_{GRB}$ = 10, 30, 100, 300 and 1000. Moreover, in order to be
free from statistical fluctuations, we have performed 10$^2$, 10$^3$, and 
10$^4$ realizations of each of these samples. 

Now the simulation of a GRB consists of the random sampling of both the 
redshift $z$ and the true $\gamma$-ray energy released $E_{\gamma} $,
according to the respective adopted distributions. Given a cosmological
model, from these coupled values we obtain the corresponding value for the 
parameter $\log d_L^2 $, which we plot on the Hubble diagram as a function 
of $z$. At this point we perform a $\chi^2 $ minimization of the simulated
data to see with what accuracy the fit reproduces the input cosmology.
The measurement error on $\log d_L^2 $ is assumed to be $\sigma = 0.3$. 

In Table~\ref{tablom} the mean results of our repeated fits are reported
for an input cosmology with $\Omega_m = 1$. 
First, from this Table it is evident that the mean values obtained from the 
fits are independent of the number of sample realizations, 
{\it i.e.}, the intrinsic fluctuations corresponding
to different samples of GRBs in the {\it same} cosmological
model, are small. Moreover, the Table shows how the accuracy of 
GRBs in constraining the matter density fraction $\Omega_m$ 
increases with their number $N_{GRB} $. The given cosmology is 
readily reproduced by the best fit value for any $N_{GRB} $, while 
its dispersion is reduced from $\sim$30\% for a sample with
$N_{GRB} =10$ to $\sim$3\% for $N_{GRB} =1000$. We have also
carried out simulations for different values of $\Omega_m$, but
still $\Omega_\Lambda = 0$, obtaining every time very similar
results.

It is worth noting that it will be very difficult and time consuming to 
determine $E_{\gamma}$ of 300 GRBs to the accuracy required. Even then the 
resulting 7\% error on $\Omega_m$ is larger than the $\sim 1\%$ errors today 
from WMAP and, eventually, SNAP. Still, an independent measurement of a 
parameter of such paramount importance need not be useless, 
even if late in coming.

\section{Simulations in a $\Lambda$-dominated cosmology}
We move now to a $\Lambda$-dominated cosmology, in which the  
contributions to the density parameter are given by the mass density, 
$\Omega_m$, and by the cosmological constant energy density, 
$\Omega_{\Lambda} $.
In the light of the recent observations of the cosmic microwave background
anisotropy (Bennett et al. 2003), we restrict our simulations to a flat 
Universe $\Omega_m +\Omega_{\Lambda} =1$. In this case the expression for the 
luminosity distance has the integral form:
\begin{equation}
d_L (z) = \frac{c}{H_0} (1+z) \int_0^z 
 \frac{dz'}{\left[ \Omega_m (1+z')^3 +\Omega_{\Lambda} \right]^{1/2} }  
\end{equation}
An analytical fit to this expression, with a relative error of less than
0.4\% for $0.2\leq \Omega_m \leq 1$, is presented in Pen (1999). In order
to reduce the run time of our simulations, we have exploited this fit to the
luminosity distance.

In a $\Lambda$-dominated cosmology, the number of GRBs per redshift unit is 
still given by equation (\ref{nzdistr}), but in this case the expression for
the comoving volume element is:
\begin{equation}
\frac{dV}{dz} = \frac{c}{H_0} \frac{4\pi d_L^2 (z)}
 {(1+z)^2 \left[\Omega_m (1+z)^3 +\Omega_{\Lambda} \right]^{1/2}}
\end{equation} 
For the GRB redshift distribution we adopt the same function as for
the $\Lambda =0$ cosmology, with the same value of $z_{max} $.
To take into account the difference in luminosity density between an
Einstein-de Sitter and a $\Lambda $ flat Universe, we applied to $n(z)$
the correction factor 
$[\Omega_m (1+z)^3 +\Omega_{\Lambda} ]^{1/2} / (1+z)^{3/2} $
(see the Appendix of Porciani \& Madau 2001 for details).

In order to study the ability of GRBs in probing the cosmological 
parameters in a $\Lambda$-dominated Universe, we have simulated 10$^2$
realizations of GRB samples with $N_{GRB}$ = 10, 30, 100, 300 and 1000.
The $\chi^2 $ minimization of the resulting Hubble diagrams has been
performed considering $\log d_L^2 $ depending only on the fit parameter
$\Omega_m $, {\it i.e.}, using the relation $\Omega_{\Lambda} = 1 -\Omega_m $.
Table~\ref{tablam} reports the general results of our repeated fits 
for a flat cosmology with input values $\Omega_m = 0.3$ and 
$\Omega_{\Lambda} = 0.7$ (which are those adopted in Frail et al. 2001).

As in the $\Lambda = 0$ case, the accuracy of GRBs in constraining the 
two contributions $\Omega_m $ and $\Omega_{\Lambda}$ to the density
parameter increases with their number $N_{GRB} $, reducing the
dispersion about the best value for the fit parameter $\Omega_m $ 
from $\sim$40\% for a sample with $N_{GRB} =10$ to $\sim$4\% for 
$N_{GRB} =1000$.

\begin{figure}
\centerline{\psfig{figure=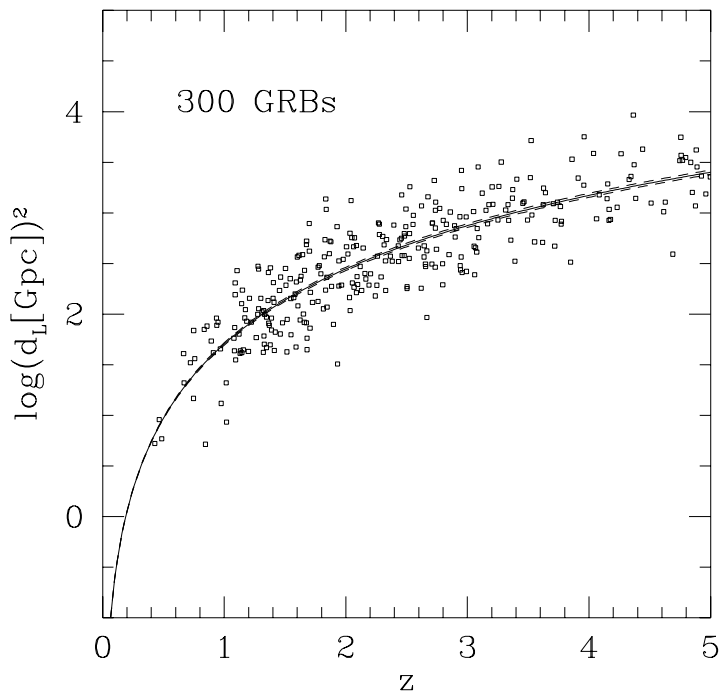,width=10cm,height=10cm}}
\caption[]{Hubble diagram $\log d_L^2 - z$ with data simulated
for a sample of 300 GRBs in a flat Universe with density parameters
$\Omega_m = 0.3$ and $\Omega_{\Lambda} = 0.7$. The solid curve shows 
the function $\log d_L^2 (z)$ in the assumed cosmology, while the dashed
curves give the dispersion about the best fit parameter (the upper curve
corresponds to lower $\Omega_m $).} 
\label{plotone}
\end{figure}

\begin{figure}
\centerline{\psfig{figure=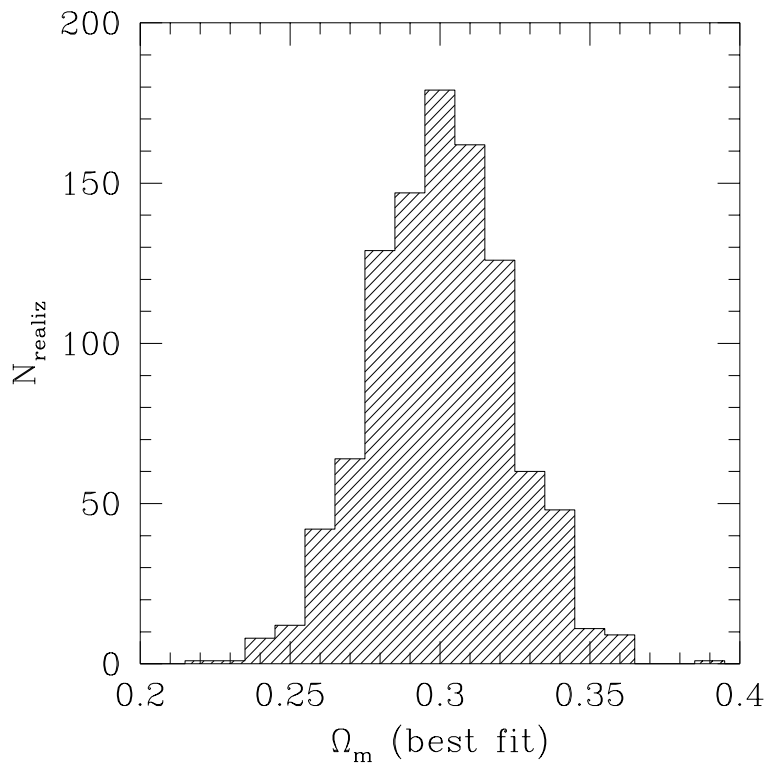,width=10cm,height=10cm}}
\caption{Histogram with the distribution of the best fit
values of the matter density $\Omega_m $ for 10$^3$ realizations of
a sample of 300 GRBs in a flat Universe with density parameters
$\Omega_m = 0.3$ and $\Omega_{\Lambda} = 0.7$. The distribution has
a mean $<\Omega_m > = 0.3001$, a median $\Omega_m (med) = 0.3002$,
a dispersion $S_{\Omega_m } = 0.0228$, and a kurtosis 
$k_{\Omega_m } = 3.0993$.}
\label{plottwo}
\end{figure}

Focussing on the samples in a $\Lambda$-dominated cosmology with 
$N_{GRB} = 300$, a data set which can be realistically obtained in
future space missions, Figure~\ref{plotone} shows one of the Hubble diagrams 
$\log d_L^2 - z$ obtained with the simulations. The solid curve shows 
the function $\log d_L^2 (z)$ in the assumed cosmology, while the dashed
curves give the dispersion about the best fit parameters. The ability of
a sample of 300 GRBs in constraining the actual cosmology is evident.
The statistical fluctuations of the $N_{GRB} = 300$ sample fit are
outlined in the histogram of Figure~\ref{plottwo}, which shows the 
distribution of the best fit values of the matter density fraction $\Omega_m $
for 10$^3$ sample realizations. The distribution peaks at 
$\Omega_m = 0.3 $, has a dispersion $S_{\Omega_m } = 0.0228$, and a
kurtosis $k_{\Omega_m } = 3.0993$, to be compared with the value of a
Gaussian distribution, {\it i.e.}, 3.

We have also investigated the effects of changing the assumptions of our
simulations on the probing ability of GRB samples to determine the geometry
of the Universe. First, we have considered a GRB redshift distribution
ruled by the simple function $n(z) \propto (1+z)^3 $ instead of 
equation~(\ref{madaufunc}). The result is a slight
decrease of the dispersion about the best fit parameter $\Omega_m $ at
all values of $N_{GRB} $. This is due to the larger number of GRBs sampled
at high redshift values by this alternative distibution, which increases
monotonically with $z$. At high redshifts the distinction between different
cosmologies becomes more evident (see curves in Figure~\ref{plotone}), 
thus more GRBs at large $z$ imply better constraints on the cosmological
parameters.  

Then we have studied the effect of varying the dispersion about the
standard candle energy. We assumed $\sigma = 0.6$ (doubling the dispersion to
a multiplicative factor of 4) about the logarithmic mean value $\mu = 51.1$
reported in Bloom, Frail, \& Kulkarni (2003). We find that the resulting 
effect is of course a worse accuracy in the reproduction of the input 
cosmology, the dispersion about the best fit parameter $\Omega_m $ 
increasing by a factor $\sim 2$ at all values of $N_{GRB} $. 
In particular, it is $\sim$15\% for a sample with $N_{GRB} =300$.

Moreover, we point out that the variation of the standard candle energy 
$E_{\gamma} $ has no effect on the ability of GRB samples in putting 
constraints on cosmological parameters, since the mean value of the
Gaussian distribution of $\log E_{\gamma} $ gives only a normalization
constant to our simulations, but is not instructive for their scatter.

Finally, we must remark that the analyses of both Frail et al. (2001) and 
Bloom, Frail, \& Kulkarni (2003) assume of course a particular set of 
cosmological parameters ($\Omega_m = 0.3$, $\Omega_{\Lambda } = 0.7 $,
and $H_0 = 65$ km s$^{-1}$ Mpc$^{-1}$) to derive the standard $\gamma$-ray 
energy of GRBs. To avoid a circular logic and the limitations in
cosmographic applications pointed out in Bloom, Frail, \& Kulkarni (2003), 
we should assume a candle calibration with a local sample of sources, a 
prospect which can now be considered possible in the light of the discovery of 
the nearby GRBs 030329 and 031203 (see Section 5 for a more detailed 
discussion on this point).

\section{Simulations in a quintessence cosmology}
\begin{figure}
\centerline{\psfig{figure=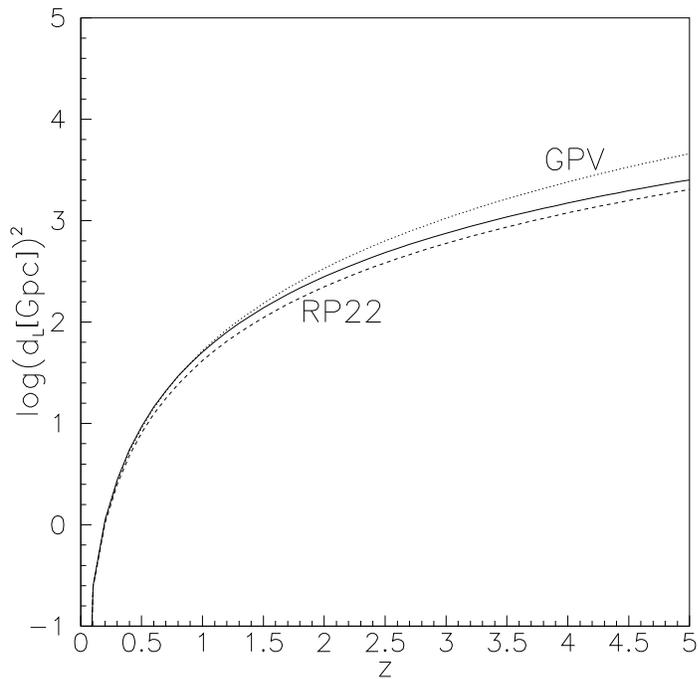,width=10cm,height=10cm}}
\caption[]{The parameter $\log d_L^2$ as a function of redshift $z$
in a flat Universe with density parameters $\Omega_m = 0.3$ and
$\Omega_{\Lambda} = 0.7$ (solid curve) compared with those obtained
in the quintessence models RP22 (lower dashed curve) and
GPV (upper dotted curve).}
\label{plotthree}
\end{figure}

Now we abandon the cosmological constant Universe and we consider some of the
most popular quintessence models. In particular, we choose as tracker
potential classes the inverse power-law Ratra-Peebles potential 
(hereafter RP; Ratra \& Peebles 1988), defined as:
\begin{equation}
V(\phi)=\frac{M^{4+\alpha}}{\phi^{\alpha}}
\end{equation}
and the SUGRA potential (Bin\'etruy 1999; Brax \& Martin 1999):
\begin{equation}
V(\phi)=\frac{M^{4+\alpha}}{\phi^{\alpha}} \exp(4\pi G\phi^2 )\;.
\end{equation}
Following Caresia, Matarrese, \& Moscardini (2004), we considered
such potentials within the framework of ``extended quintessence''
models. These are characterized by a coupling between gravity and
quintessence ruled by a parameter $\xi$, where $\xi =0$ means no
coupling. By fixing $\alpha$ and $\xi$, we obtained numerical
values for the luminosity distance $d_L (z)$ with the standard
procedure. First, we numerically solved the Klein-Gordon equation
for the scalar field $\phi$. From this solution it is possible to
get the expansion rate $H(z)=H_0 h(z)$, which depends on $\phi$
since the quintessence scalar field contributes to the total
energy density. Then we exploited the usual relation between
expansion rate and luminosity distance:
\begin{equation}
d_L (z) = \frac{c}{H_0} (1+z) \int_0^z \frac{dz'}{h(z')}
\end{equation}
We repeated the same procedure for the dilaton scenario introduced
by Gasperini, Piazza, \& Veneziano (2002; hereafter GPV). In order
to distinguish between different tracker potentials of the same
class, say RP, we will use a couple of indices, the first one
giving the value of the $\alpha$ parameter, while the second one
referring to the adopted coupling parameter $\xi$, being 1 for
$\xi=0$ and 2 for $\xi=0.01$ (therefore, model RP01 corresponds to
no quintessence). In Figure~\ref{plotthree} we report
the function $\log d_L^2 (z)$ found for the two most ``extreme''
among the quintessence models considered, {\it i.e.}, those in
which this function differs most strongly from that corresponding
to a $\Lambda$-dominated flat Universe with density parameters
$\Omega_m = 0.3$ and $\Omega_{\Lambda} = 0.7$, already shown in
Figure~\ref{plotone}. All the other models give a $\log d_L^2 (z)$
curve lying between the no quintessence case and the RP22 curve.{}

Figure~\ref{plotthree} allows us to make an important, technical
point. We have so far postponed the thorny issue of the
calibration of the absolute energy release: in fact, when a sample
of GRBs is used to derive the distribution of energy releases, it
is necessary to assume a set of cosmological parameters to compare
GRBs at different redshifts, making (potentially!) our argument
circular. However, it is well known that, for $z\ll 1$, all
cosmological models coincide, so that a {\it local} calibration of
the absolute energy release is possible. Originally, given the
very large redshifts of the first GRBs, it was not clear whether
this could be achieved, but with the accumulation of further data
this does not look like a real concern: there are currently 4
GRBs with $z < 0.4$, out of a total of 39 (Figure~\ref{plotfour}), making a 
{\it local} calibration with a large sample of GRBs a real possibility.

Furthermore, the situation is even better if we assume
$\Omega_m +\Omega_\Lambda = 1$: we see from Figure~\ref{plotthree}
that all models, including GPV which is by far the most discordant
one, yield essentially the same luminosity distance out to
$z\approx 1$. There are currently 17 GRBs with z $< 1$
out of a total of 39 (Figure~\ref{plotfour}), making the issue of calibration a
moot one, once a sufficiently large sample is obtained.

In order to determine the number of GRBs per redshift unit in quintessence
universes, we can exploit again the Appendix of Porciani \& Madau (2001) and
find that:
\begin{equation}
\frac{dN_{GRB}}{dz} = \frac{d_L^2(z)n(z)/(1+z)^{7/2}}
{\int_0^{z_{max}} d_L^2(z)n(z)/(1+z)^{7/2} dz}
\label{nzdistrgen}
\end{equation}
independently of $dV/dz$, where we have already taken into account the 
correction factor for the difference in luminosity density from the 
Einstein-de Sitter Universe. The adopted GRB redshift distribution $n(z)$ 
is always given by equation~(\ref{madaufunc}), and we fix again $z_{max} =5$.

Our purpose is now to investigate whether it is possible to
discriminate between different quintessence cosmological models
via a set of GRBs, considered as standard candles. To do this, we
performed a series of KS tests on simulated data sets with
$N_{GRB}$ =100, 300 and 1000 in two different $\Lambda$-dominated
flat cosmologies both with $\Omega_m = 0.3$ and $\Omega_{\Lambda }
= 0.7$ at $z=0$, but, while one has a truly constant $\Lambda$, the other one 
has a quintessence field which reduces to
$\Omega_\Lambda =0.7$ at $z=0$. The model for quintessence is
chosen from those listed before. We considered both the case in
which the same set of GRB redshifts is used, and the 
case in which different patches of the Universe are
selected, each with its own cosmology, and thus two fully distinct
sets of GRB redshifts are used for the two different cosmologies.
In this second way, we wish to include the cosmic variance into
our simulations. Our results are presented in
Table~\ref{tabquint}, for a same GRB redshift distribution, and in
Table~\ref{tabquintvar}, where cosmic variance has been
considered.

From Tables~\ref{tabquint} and \ref{tabquintvar} we see that it is quite
difficult to discriminate a quintessence cosmological model using a set of up
to 1000 GRBs as standard candles. Only the GPV model could be significantly
discriminated with a set of 1000 GRBs, especially if we take into account
cosmic variance, in which case we obtained $<Q_{KS}>=4.89\cdot 10^{-8}$.
A discrimination may also be possible for the RP22 model, which resulted in
a $<Q_{KS}>=8.45\cdot 10^{-3}$ with 1000 GRBs, when not considering cosmic
variance. These results are consistent with the extreme behaviours of the
$\log d_L^2 (z)$ functions for GPV and RP22 models, as already
stated before and shown in Figure~\ref{plotthree}.

These findings may be the consequence of the high dispersion,
$\sigma =0.3$ in its logarithm, around the mean value of the
geometrically corrected $\gamma$-ray energy released by GRBs, which we
have taken as a cosmological candle. In order to investigate whether a
better discrimination of different quintessence models could be
obtained with a pin down of our standard candle, we performed
another series of KS tests on simulated data sets, in which we
adopted a SNIa-like candle with a logarithmic dispersion of
$\sigma =0.072$, corresponding to a conservative magnitude
dispersion of $\sigma_M =0.18$ mag, intermediate between the
$\sigma_M =0.21$ mag value found by Riess, Press, \& Kirshner
(1995) and the later $\sigma_M =0.12$ mag of the same authors
(Riess, Press, \& Kirshner 1996) together with the very recent
results by Wang et al. (2003) in the range $\sigma_M =0.08-0.11$
mag. The conditions of the simulations were the same as for the
first KS test series, with the additional subdivision of a value
$z_{max}=5$, typical of GRBs, or 2, more typical of SNIa. In
the first case, our results are reported in Table~\ref{tabsnIa5},
for the same standard candle redshift distribution, and in
Table~\ref{tabsnIa5var}, where cosmic variance has been
considered, while in the second case the results are shown
respectively in Table~\ref{tabsnIa2} and Table~\ref{tabsnIa2var}.

Analysis of Tables~\ref{tabsnIa5}$-$\ref{tabsnIa2var} shows that
with a less dispersed cosmological candle the GPV and RP22
quintessence models could be significantly discriminated with a
set of 1000 sources observed up to $z_{max}=5$, the first one in
this case perhaps even with only 300 sources ($<Q_{KS}>~\! =1.04\cdot 10^{-3}$ 
taking into account cosmic variance). Moreover, a hint
for discrimination is also given with 1000 candles by the RP21
model, which resulted in a $<Q_{KS}>=6.92\cdot 10^{-3}$ when not
considering cosmic variance. On the other hand, limiting the
candle distribution to $z_{max}=2$, it seems possible to
significantly discriminate all the quintessence models of the RP
tracker potential class, together with the GPV model, if 1000
sources were observed. The RP21 and RP22 models may be
discriminated even with a data set of only 300 standard candles.
These results are consistent with Figure~\ref{plotthree}, where it
is possible to notice how up to $z=2$ the $\log d_L^2 (z)$
function for the RP22 (and also for the not reported RP21) model
differs more than the GPV one from the no quintessence case. There 
seems to be no chance for a discrimination of the SUGRA class of
quintessence models, whose $\log d_L^2 (z)$ functions are in fact
very similar to the solide curve of Figure~\ref{plotthree}.

\section{Conclusions}
We have simulated different samples of
GRBs adopting $\gamma$-ray energy and redshift distributions consistent
with recent observational results, in order to investigate their ability
to probe cosmological parameters such as the density fractions  
$\Omega_m $ and $\Omega_{\Lambda }$. Our result is that in a
$\Lambda $-dominated flat Universe the accuracy
in the determination of the matter density $\Omega_m $ is  
$\sim$40\% for a sample with $N_{GRB} =10$ and an excellent $\sim$4\% for 
$N_{GRB} =1000$. 

For comparison, a $\sim$20\% accuracy on the determination of $\Omega_m $ has 
been recently claimed by using GRBs as standard candles ruled by the 
luminosity-variability and luminosity-lag time relations 
(Takahashi et al. 2003).
 
\begin{figure}
\centerline{\psfig{figure=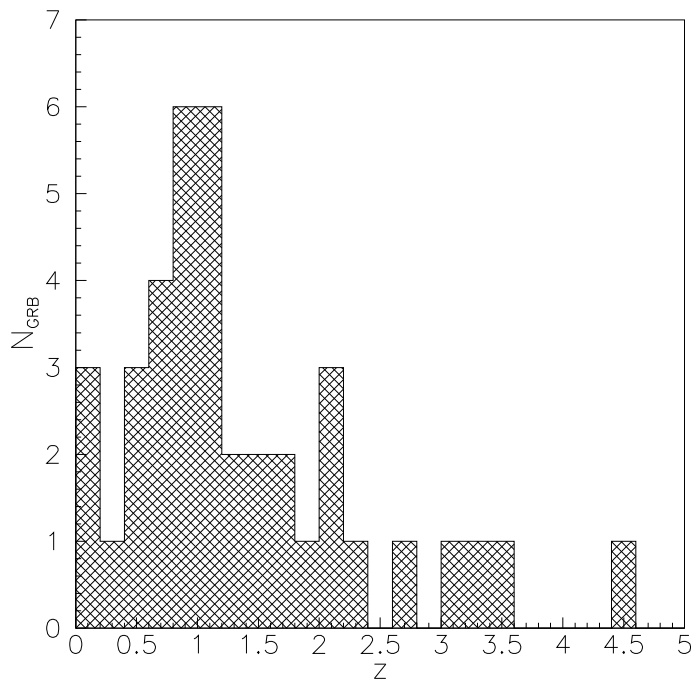,width=10cm,height=10cm}}
\caption{Distribution of all the known redshifts of GRBs as of November 2004.
All redshifts are within the 0.1$-$4.5 range, apart from GRB 980425,
possibly associated with the nearby SN1998bw (z=0.0085, Galama et al. 1998).
Data is taken from Greiner (2004).}
\label{plotfour}
\end{figure}

Since GRBs are much more readily observed than SNIa, 
especially at high redshifts (notice that in the sample
observed so far, which is reported in Figure~\ref{plotfour}, there
are already 4 GRBs with $z >3$, about 10\% of the whole), they
should allow us to probe cosmological parameters more deeply than
these latter sources.
Moreover, during the last few years of observations the number of GRBs with 
known redshifts has almost reached the same number of high redshift SNIa 
discovered by the Supernova Cosmology Project (Perlmutter et al. 1999).

Lastly, after showing that the absolute energy release can be
calibrated using the low redshift GRBs, we have shown that GRBs
have the potential to investigate the luminosity distance out to
large redshifts, and this, in turn, means that at least some
models for quintessence, among which the important dilaton model
of Gasperini, Piazza and Veneziano, can be tested and discriminated from
competing models.

On 20 November 2004, the {\it Swift} satellite was launched, and
the detection of $\sim$200 GRBs with known redshifts is expected
during the 3 years of its sky observations. If GRBs are confirmed by
these new data to emit a standard amount of energy, then our
simulations stress how the accuracy in the determination of
cosmological parameters increases with the number of their known
redshifts, making {\it Swift}, at least potentially, a GRB
Cosmology Project.

\newpage

\section*{References}
\begin{harvard}
\item[] Bennett C L et al. 2003 {\it ApJS} {\bf 148} 1
\item[] Berger E, Kulkarni S R and Frail D A 2003 {\it ApJ} {\bf 590} 379 
\item[] Bin\'etruy P 1999 {\it Phys. Rev.} D {\bf 60} 063502
\item[] Bloom J S, Frail D A and Kulkarni S R 2003 {\it ApJ} {\bf 594} 674
\item[] Bloom J S, Frail D A and Sari R 2001 {\it AJ} {\bf 121} 2879 
\item[] Brax P and Martin J 1999 {\it Phys. Lett.} B {\bf 468} 40
\item[] Caresia P, Matarrese S and Moscardini L 2004 {\it ApJ} {\bf 605} 21
\item[] Cohen E and Piran T 1997 {\it ApJ} {\bf 488} L7
\item[] Dai Z G, Liang E W and Xu D 2004 {\it ApJ} {\bf 612} L101 
\item[] Frail D A et al. 2001 {\it ApJ} {\bf 562} L55 
\item[] Galama T J et al. 1998 {\it Nature} {\bf 395} 670
\item[] Gasperini M, Piazza F and Veneziano G 2002 {\it Phys. Rev.} D {\bf 65}
023508 (GPV)
\item[] Ghirlanda G, Ghisellini G and Lazzati D 2004 {\it ApJ} {\bf 616} 331
\item[] Ghirlanda G, Ghisellini G, Lazzati D and Firmani C 2004 {\it ApJ} 
{\bf 613} L13
\item[] Greiner J 2004 GRB webpage, http://www.mpe.mpg.de/$\sim$jcg/grbgen.html
\item[] Malesani D et al. 2004 {\it ApJ} {\bf 609} L5
\item[] Panaitescu A and Kumar P 2001 {\it ApJ} {\bf 560} L49 
\item[] Pen U L 1999 {\it ApJS} {\bf 120} 49
\item[] Perlmutter S et al. 1999 {\it ApJ} {\bf 517} 565
\item[] Piran T et al. 2001 {\it ApJ} {\bf 560} L167 
\item[] Porciani C and Madau P 2001 {\it ApJ} {\bf 548} 522 
\item[] Price P A et al. 2003 {\it Nature} {\bf 423} 844
\item[] Ratra B and Peebles P J E 1988 {\it Phys. Rev.} D {\bf 37} 3406 (RP)
\item[] Riess A G, Press W H and Kirshner R P 1995 {\it ApJ} {\bf 438} L17
\item[] \dash 1996 {\it ApJ} {\bf 473} 88
\item[] Schaefer B E 2003 {\it ApJ} {\bf 583} L67
\item[] Takahashi K, Oguri M, Kotake K and Ohno H 2003 {\it Preprint} 
astro-ph/0305260
\item[] Von Mises R 1964 {\it Mathematical Theory of Probability and 
Statistics} (New York: Academic Press) Chapter~IX
\item[] Wang L, Goldhaber G, Aldering G and Perlmutter S 2003 {\it ApJ} 
{\bf 590} 944
\end{harvard}

\newpage

\begin{table}
\caption{\label{tablom}Mean values of the fitted cosmological density 
parameter $\Omega_m$, of its error $\Delta \Omega_m$ and its dispersion 
$S_{\Omega_m }$ obtained by fitting 10$^2$ (top), 10$^3$ (middle) 
and 10$^4$ (bottom) GRB sample realizations with $N_{GRB} $ distributed 
according to function $R_{SF1}(z)$ of Porciani \& Madau (2001) in an 
Einstein-de Sitter Universe ($\Omega_m = 1$).}
\begin{indented}
\item[]\begin{tabular}{@{}cccc}
\br
$N_{GRB} $ & $<\Omega_m >$ & $<\Delta \Omega_m >$ & $S_{\Omega_m}$ \\
\mr
 10    & 0.9983 & 0.2997 & 0.3097 \\
 30    & 1.0158 & 0.1895 & 0.1993 \\ 
 100   & 0.9937 & 0.0993 & 0.1108 \\ 
 300   & 0.9959 & 0.0599 & 0.0629 \\
 1000  & 1.0009 & 0.0332 & 0.0351 \\
\br
\end{tabular}

\vspace{3mm}
\item[]\begin{tabular}{@{}cccc}
\br
$N_{GRB} $ & $<\Omega_m >$ & $<\Delta \Omega_m >$ & $S_{\Omega_m}$ \\
\mr
 10    & 1.0265 & 0.3095 & 0.3483 \\
 30    & 1.0142 & 0.1766 & 0.1993 \\ 
 100   & 1.0026 & 0.0993 & 0.1048 \\ 
 300   & 1.0020 & 0.0593 & 0.0595 \\
 1000  & 1.0015 & 0.0331 & 0.0333 \\
\br
\end{tabular}

\vspace{3mm}
\item[]\begin{tabular}{@{}cccc}
\br
$N_{GRB} $ & $<\Omega_m >$ & $<\Delta \Omega_m >$ & $S_{\Omega_m}$ \\
\mr
 10    & 1.0320 & 0.3085 & 0.3516 \\
 30    & 1.0103 & 0.1780 & 0.1939 \\ 
 100   & 1.0043 & 0.0997 & 0.1074 \\ 
 300   & 1.0011 & 0.0590 & 0.0614 \\
 1000  & 1.0004 & 0.0330 & 0.0332 \\
\br
\end{tabular}
\end{indented}
\end{table}

\begin{table}
\caption{\label{tablam} Mean values of the fitted cosmological density 
parameters $\Omega_m$ and $\Omega_{\Lambda} $, 
of their error $\Delta \Omega$ and their dispersion 
$S_{\Omega}$ obtained by fitting 10$^2$ GRB sample realizations 
with $N_{GRB} $ distributed according to function $R_{SF1}(z)$ of 
Porciani \& Madau (2001) in a flat Universe with input values 
$\Omega_m = 0.3$ and $\Omega_{\Lambda} = 0.7$.}
\begin{indented}
\item[]\begin{tabular}{@{}ccccc}
\br
$N_{GRB} $ & $<\Omega_m >$ & $<\Omega_{\Lambda} >$ &
$<\Delta \Omega >$ & $S_{\Omega} $ \\
\mr
 10    & 0.3195 & 0.6805 & 0.1004 & 0.1307 \\
 30    & 0.2973 & 0.7027 & 0.0763 & 0.0700 \\ 
 100   & 0.3002 & 0.6998 & 0.0363 & 0.0351 \\ 
 300   & 0.3023 & 0.6977 & 0.0219 & 0.0222 \\
 1000  & 0.3001 & 0.6999 & 0.0120 & 0.0125 \\
\br
\end{tabular}
\end{indented}
\end{table}

\begin{table}
\caption{\label{tabquint}Mean values of the probability $Q_{KS}$ and the 
maximum distance $D$ for a KS test on the parameter $\log d_L^2 (z)$ of
100 realizations of a redshift sample made of 100 (top), 300 (middle)
and 1000 (bottom) GRBs obtained in a $\Lambda$-dominated flat cosmology
with $\Omega_m =0.3$ and $\Omega_{\Lambda} =0.7$ at $z=0$, but with two 
different $\log d_L^2 (z)$ distributions, one resulting from a cosmology with
a truly constant $\Lambda$ and the other from a quintessence model defined by
its name in the first column, which reduces to the same cosmology at $z=0$.}
\begin{indented}
\item[]\begin{tabular}{@{}ccc}
\br
  Model  &       $<Q_{KS}>$       &   $<D>$  \\
\mr
 RP01    &   $6.84\cdot 10^{-1}$  &   0.098  \\
 RP11    &   $6.08\cdot 10^{-1}$  &   0.107  \\
 RP12    &   $5.75\cdot 10^{-1}$  &   0.110  \\
 RP21    &   $4.91\cdot 10^{-1}$  &   0.120  \\
 RP22    &   $4.54\cdot 10^{-1}$  &   0.125  \\
 SUGRA11 &   $7.29\cdot 10^{-1}$  &   0.094  \\
 SUGRA12 &   $7.25\cdot 10^{-1}$  &   0.094  \\
 SUGRA21 &   $6.37\cdot 10^{-1}$  &   0.104  \\
 SUGRA22 &   $6.52\cdot 10^{-1}$  &   0.103  \\
 GPV     &   $2.76\cdot 10^{-1}$  &   0.149  \\
\br
\end{tabular}

\vspace{3mm}
\item[]\begin{tabular}{@{}ccc}
\br
  Model  &       $<Q_{KS}>$       &   $<D>$  \\
\mr
 RP01    &   $6.98\cdot 10^{-1}$  &   0.057  \\
 RP11    &   $4.37\cdot 10^{-1}$  &   0.073  \\
 RP12    &   $4.22\cdot 10^{-1}$  &   0.076  \\
 RP21    &   $2.14\cdot 10^{-1}$  &   0.094  \\
 RP22    &   $1.41\cdot 10^{-1}$  &   0.102  \\
 SUGRA11 &   $6.38\cdot 10^{-1}$  &   0.060  \\
 SUGRA12 &   $6.12\cdot 10^{-1}$  &   0.061  \\
 SUGRA21 &   $5.05\cdot 10^{-1}$  &   0.069  \\
 SUGRA22 &   $5.26\cdot 10^{-1}$  &   0.067  \\
 GPV     &   $8.32\cdot 10^{-2}$  &   0.113  \\
\br
\end{tabular}

\vspace{3mm}
\item[]\begin{tabular}{@{}ccc}
\br
  Model  &       $<Q_{KS}>$       &   $<D>$  \\
\mr
 RP01    &   $7.16\cdot 10^{-1}$  &   0.031  \\
 RP11    &   $1.85\cdot 10^{-1}$  &   0.053  \\
 RP12    &   $1.35\cdot 10^{-1}$  &   0.056  \\
 RP21    &   $1.98\cdot 10^{-2}$  &   0.076  \\
 RP22    &   $8.45\cdot 10^{-3}$  &   0.082  \\
 SUGRA11 &   $4.70\cdot 10^{-1}$  &   0.039  \\
 SUGRA12 &   $4.56\cdot 10^{-1}$  &   0.040  \\
 SUGRA21 &   $3.19\cdot 10^{-1}$  &   0.045  \\
 SUGRA22 &   $2.94\cdot 10^{-1}$  &   0.047  \\
 GPV     &   $3.28\cdot 10^{-4}$  &   0.103  \\
\br
\end{tabular}
\end{indented}
\end{table}

\begin{table}
\caption{\label{tabquintvar}Mean values of the probability $Q_{KS}$ and the 
maximum distance $D$ for a KS test on the parameter $\log d_L^2 (z)$ of
100 realizations of two different samples made of 100 (top), 300 (middle)
and 1000 (bottom) GRBs, one obtained in a $\Lambda$-dominated flat cosmology
with $\Omega_m =0.3$ and $\Omega_{\Lambda} =0.7$, and the other in
a quintessence model defined by its name in the 
first column, which reduces to the same cosmology at $z=0$.}
\begin{indented}
\item[]\begin{tabular}{@{}ccc}
\br
  Model  &       $<Q_{KS}>$       &   $<D>$  \\
\mr
 RP01    &   $4.64\cdot 10^{-1}$  &   0.126  \\
 RP11    &   $4.91\cdot 10^{-1}$  &   0.123  \\
 RP12    &   $4.91\cdot 10^{-1}$  &   0.123  \\
 RP21    &   $4.60\cdot 10^{-1}$  &   0.128  \\
 RP22    &   $3.55\cdot 10^{-1}$  &   0.143  \\
 SUGRA11 &   $4.97\cdot 10^{-1}$  &   0.121  \\
 SUGRA12 &   $4.92\cdot 10^{-1}$  &   0.121  \\
 SUGRA21 &   $4.52\cdot 10^{-1}$  &   0.127  \\
 SUGRA22 &   $5.42\cdot 10^{-1}$  &   0.117  \\
 GPV     &   $1.41\cdot 10^{-1}$  &   0.190  \\
\br
\end{tabular}

\vspace{3mm}
\item[]\begin{tabular}{@{}ccc}
\br
  Model  &       $<Q_{KS}>$       &   $<D>$  \\
\mr
 RP01    &   $5.01\cdot 10^{-1}$  &   0.070  \\
 RP11    &   $3.45\cdot 10^{-1}$  &   0.082  \\
 RP12    &   $4.16\cdot 10^{-1}$  &   0.076  \\
 RP21    &   $2.30\cdot 10^{-1}$  &   0.099  \\
 RP22    &   $1.80\cdot 10^{-1}$  &   0.106  \\
 SUGRA11 &   $5.16\cdot 10^{-1}$  &   0.069  \\
 SUGRA12 &   $5.06\cdot 10^{-1}$  &   0.070  \\
 SUGRA21 &   $4.18\cdot 10^{-1}$  &   0.077  \\
 SUGRA22 &   $4.50\cdot 10^{-1}$  &   0.074  \\
 GPV     &   $7.36\cdot 10^{-3}$  &   0.173  \\
\br
\end{tabular}

\vspace{3mm}
\item[]\begin{tabular}{@{}ccc}
\br
  Model  &       $<Q_{KS}>$       &   $<D>$  \\
\mr
 RP01    &   $5.31\cdot 10^{-1}$  &   0.037  \\
 RP11    &   $2.32\cdot 10^{-1}$  &   0.053  \\
 RP12    &   $1.49\cdot 10^{-1}$  &   0.062  \\
 RP21    &   $5.43\cdot 10^{-2}$  &   0.077  \\
 RP22    &   $1.81\cdot 10^{-2}$  &   0.086  \\
 SUGRA11 &   $3.74\cdot 10^{-1}$  &   0.045  \\
 SUGRA12 &   $4.09\cdot 10^{-1}$  &   0.043  \\
 SUGRA21 &   $3.54\cdot 10^{-1}$  &   0.047  \\
 SUGRA22 &   $2.66\cdot 10^{-1}$  &   0.051  \\
 GPV     &   $4.89\cdot 10^{-8}$  &   0.158  \\
\br
\end{tabular}
\end{indented}
\end{table}

\begin{table}
\caption{\label{tabsnIa5} The same as Table~\ref{tabquint} but obtained by 
adopting SNIa-like standard candles distributed with a logarithmic dispersion
$\sigma =0.072$ up to $z_{max}=5$.}
\begin{indented}
\item[]\begin{tabular}{@{}ccc}
\br
  Model  &       $<Q_{KS}>$       &   $<D>$  \\
\mr
 RP01    &   $9.66\cdot 10^{-1}$  &   0.062  \\
 RP11    &   $7.83\cdot 10^{-1}$  &   0.090  \\
 RP12    &   $7.20\cdot 10^{-1}$  &   0.096  \\
 RP21    &   $5.45\cdot 10^{-1}$  &   0.112  \\
 RP22    &   $4.27\cdot 10^{-1}$  &   0.125  \\
 SUGRA11 &   $9.21\cdot 10^{-1}$  &   0.070  \\
 SUGRA12 &   $8.99\cdot 10^{-1}$  &   0.075  \\
 SUGRA21 &   $8.52\cdot 10^{-1}$  &   0.081  \\
 SUGRA22 &   $8.21\cdot 10^{-1}$  &   0.086  \\
 GPV     &   $1.93\cdot 10^{-1}$  &   0.157  \\
\br
\end{tabular}

\vspace{3mm}
\item[]\begin{tabular}{@{}ccc}
\br
  Model  &       $<Q_{KS}>$       &   $<D>$  \\
\mr
 RP01    &   $9.68\cdot 10^{-1}$  &   0.037  \\
 RP11    &   $5.11\cdot 10^{-1}$  &   0.067  \\
 RP12    &   $4.45\cdot 10^{-1}$  &   0.071  \\
 RP21    &   $1.65\cdot 10^{-1}$  &   0.092  \\
 RP22    &   $1.09\cdot 10^{-1}$  &   0.099  \\
 SUGRA11 &   $8.19\cdot 10^{-1}$  &   0.050  \\
 SUGRA12 &   $8.18\cdot 10^{-1}$  &   0.050  \\
 SUGRA21 &   $6.59\cdot 10^{-1}$  &   0.059  \\
 SUGRA22 &   $6.39\cdot 10^{-1}$  &   0.060  \\
 GPV     &   $2.03\cdot 10^{-2}$  &   0.129  \\
\br
\end{tabular}

\vspace{3mm}
\item[]\begin{tabular}{@{}ccc}
\br
  Model  &       $<Q_{KS}>$       &   $<D>$  \\
\mr
 RP01    &   $9.68\cdot 10^{-1}$  &   0.020  \\
 RP11    &   $1.39\cdot 10^{-1}$  &   0.052  \\
 RP12    &   $9.51\cdot 10^{-2}$  &   0.056  \\
 RP21    &   $6.92\cdot 10^{-3}$  &   0.076  \\
 RP22    &   $1.54\cdot 10^{-3}$  &   0.087  \\
 SUGRA11 &   $5.49\cdot 10^{-1}$  &   0.036  \\
 SUGRA12 &   $5.17\cdot 10^{-1}$  &   0.037  \\
 SUGRA21 &   $2.95\cdot 10^{-1}$  &   0.044  \\
 SUGRA22 &   $2.45\cdot 10^{-1}$  &   0.046  \\
 GPV     &   $5.57\cdot 10^{-6}$  &   0.118  \\
\br
\end{tabular}
\end{indented}
\end{table}

\begin{table}
\caption{\label{tabsnIa5var} The same as Table~\ref{tabquintvar} but obtained 
by adopting SNIa-like standard candles distributed with a logarithmic 
dispersion $\sigma =0.072$ up to $z_{max}=5$.}
\begin{indented}
\item[]\begin{tabular}{@{}ccc}
\br
  Model  &       $<Q_{KS}>$       &   $<D>$  \\
\mr
 RP01    &   $5.25\cdot 10^{-1}$  &   0.118  \\
 RP11    &   $4.81\cdot 10^{-1}$  &   0.122  \\
 RP12    &   $4.55\cdot 10^{-1}$  &   0.127  \\
 RP21    &   $4.06\cdot 10^{-1}$  &   0.138  \\
 RP22    &   $3.36\cdot 10^{-1}$  &   0.145  \\
 SUGRA11 &   $4.59\cdot 10^{-1}$  &   0.126  \\
 SUGRA12 &   $4.81\cdot 10^{-1}$  &   0.121  \\
 SUGRA21 &   $4.66\cdot 10^{-1}$  &   0.125  \\
 SUGRA22 &   $4.68\cdot 10^{-1}$  &   0.128  \\
 GPV     &   $6.28\cdot 10^{-2}$  &   0.217  \\
\br
\end{tabular}

\vspace{3mm}
\item[]\begin{tabular}{@{}ccc}
\br
  Model  &       $<Q_{KS}>$       &   $<D>$  \\
\mr
 RP01    &   $5.07\cdot 10^{-1}$  &   0.069  \\
 RP11    &   $3.56\cdot 10^{-1}$  &   0.083  \\
 RP12    &   $3.21\cdot 10^{-1}$  &   0.088  \\
 RP21    &   $1.97\cdot 10^{-1}$  &   0.101  \\
 RP22    &   $1.09\cdot 10^{-1}$  &   0.118  \\
 SUGRA11 &   $4.52\cdot 10^{-1}$  &   0.075  \\
 SUGRA12 &   $3.92\cdot 10^{-1}$  &   0.079  \\
 SUGRA21 &   $4.14\cdot 10^{-1}$  &   0.078  \\
 SUGRA22 &   $3.38\cdot 10^{-1}$  &   0.083  \\
 GPV     &   $1.04\cdot 10^{-3}$  &   0.191  \\
\br
\end{tabular}

\vspace{3mm}
\item[]\begin{tabular}{@{}ccc}
\br
  Model  &       $<Q_{KS}>$       &   $<D>$  \\
\mr
 RP01    &   $5.22\cdot 10^{-1}$  &   0.038  \\
 RP11    &   $1.57\cdot 10^{-1}$  &   0.059  \\
 RP12    &   $1.27\cdot 10^{-1}$  &   0.063  \\
 RP21    &   $1.89\cdot 10^{-2}$  &   0.082  \\
 RP22    &   $1.58\cdot 10^{-2}$  &   0.089  \\
 SUGRA11 &   $3.58\cdot 10^{-1}$  &   0.045  \\
 SUGRA12 &   $3.62\cdot 10^{-1}$  &   0.046  \\
 SUGRA21 &   $2.27\cdot 10^{-1}$  &   0.053  \\
 SUGRA22 &   $2.43\cdot 10^{-1}$  &   0.051  \\
 GPV     &   $6.07\cdot 10^{-10}$ &   0.177  \\
\br
\end{tabular}
\end{indented}
\end{table}

\begin{table}
\caption{\label{tabsnIa2} The same as Table~\ref{tabquint} but obtained by 
adopting SNIa-like standard candles distributed with a logarithmic dispersion
$\sigma =0.072$ up to $z_{max}=2$.}
\begin{indented}
\item[]\begin{tabular}{@{}ccc}
\br
  Model  &       $<Q_{KS}>$       &   $<D>$  \\
\mr
 RP01    &   $8.78\cdot 10^{-1}$  &   0.077  \\
 RP11    &   $3.75\cdot 10^{-1}$  &   0.131  \\
 RP12    &   $2.99\cdot 10^{-1}$  &   0.141  \\
 RP21    &   $8.10\cdot 10^{-2}$  &   0.188  \\
 RP22    &   $6.73\cdot 10^{-2}$  &   0.192  \\
 SUGRA11 &   $7.14\cdot 10^{-1}$  &   0.096  \\
 SUGRA12 &   $6.61\cdot 10^{-1}$  &   0.101  \\
 SUGRA21 &   $5.55\cdot 10^{-1}$  &   0.113  \\
 SUGRA22 &   $5.12\cdot 10^{-1}$  &   0.116  \\
 GPV     &   $3.33\cdot 10^{-1}$  &   0.140  \\
\br
\end{tabular}

\vspace{3mm}
\item[]\begin{tabular}{@{}ccc}
\br
  Model  &       $<Q_{KS}>$       &   $<D>$  \\
\mr
 RP01    &   $8.89\cdot 10^{-1}$  &   0.044  \\
 RP11    &   $7.51\cdot 10^{-2}$  &   0.108  \\
 RP12    &   $5.24\cdot 10^{-2}$  &   0.116  \\
 RP21    &   $2.32\cdot 10^{-3}$  &   0.157  \\
 RP22    &   $5.57\cdot 10^{-4}$  &   0.176  \\
 SUGRA11 &   $4.56\cdot 10^{-1}$  &   0.071  \\
 SUGRA12 &   $4.33\cdot 10^{-1}$  &   0.072  \\
 SUGRA21 &   $2.19\cdot 10^{-1}$  &   0.089  \\
 SUGRA22 &   $1.80\cdot 10^{-1}$  &   0.092  \\
 GPV     &   $4.93\cdot 10^{-2}$  &   0.116  \\
\br
\end{tabular}

\vspace{3mm}
\item[]\begin{tabular}{@{}ccc}
\br
  Model  &       $<Q_{KS}>$       &   $<D>$  \\
\mr
 RP01    &   $8.77\cdot 10^{-1}$  &   0.025  \\
 RP11    &   $1.03\cdot 10^{-3}$  &   0.094  \\
 RP12    &   $2.90\cdot 10^{-4}$  &   0.100  \\
 RP21    &   $5.57\cdot 10^{-8}$  &   0.143  \\
 RP22    &   $1.53\cdot 10^{-9}$  &   0.158  \\
 SUGRA11 &   $1.10\cdot 10^{-1}$  &   0.057  \\
 SUGRA12 &   $1.04\cdot 10^{-1}$  &   0.057  \\
 SUGRA21 &   $1.85\cdot 10^{-2}$  &   0.072  \\
 SUGRA22 &   $1.22\cdot 10^{-2}$  &   0.076  \\
 GPV     &   $1.25\cdot 10^{-4}$  &   0.104  \\
\br
\end{tabular}
\end{indented}
\end{table}

\begin{table}
\caption{\label{tabsnIa2var} The same as Table~\ref{tabquintvar} but obtained 
by adopting SNIa-like standard candles distributed with a logarithmic 
dispersion $\sigma =0.072$ up to $z_{max}=2$.}
\begin{indented}
\item[]\begin{tabular}{@{}ccc}
\br
  Model  &       $<Q_{KS}>$       &   $<D>$  \\
\mr
 RP01    &   $4.94\cdot 10^{-1}$  &   0.122  \\
 RP11    &   $2.89\cdot 10^{-1}$  &   0.153  \\
 RP12    &   $3.09\cdot 10^{-1}$  &   0.150  \\
 RP21    &   $1.50\cdot 10^{-1}$  &   0.189  \\
 RP22    &   $1.01\cdot 10^{-1}$  &   0.199  \\
 SUGRA11 &   $4.53\cdot 10^{-1}$  &   0.128  \\
 SUGRA12 &   $3.92\cdot 10^{-1}$  &   0.135  \\
 SUGRA21 &   $3.82\cdot 10^{-1}$  &   0.139  \\
 SUGRA22 &   $3.52\cdot 10^{-1}$  &   0.142  \\
 GPV     &   $2.42\cdot 10^{-1}$  &   0.166  \\
\br
\end{tabular}

\vspace{3mm}
\item[]\begin{tabular}{@{}ccc}
\br
  Model  &       $<Q_{KS}>$       &   $<D>$  \\
\mr
 RP01    &   $4.81\cdot 10^{-1}$  &   0.070  \\
 RP11    &   $1.23\cdot 10^{-1}$  &   0.116  \\
 RP12    &   $8.68\cdot 10^{-2}$  &   0.117  \\
 RP21    &   $8.56\cdot 10^{-3}$  &   0.162  \\
 RP22    &   $2.67\cdot 10^{-3}$  &   0.179  \\
 SUGRA11 &   $3.27\cdot 10^{-1}$  &   0.084  \\
 SUGRA12 &   $3.41\cdot 10^{-1}$  &   0.084  \\
 SUGRA21 &   $2.30\cdot 10^{-1}$  &   0.097  \\
 SUGRA22 &   $2.19\cdot 10^{-1}$  &   0.098  \\
 GPV     &   $3.89\cdot 10^{-2}$  &   0.138  \\
\br
\end{tabular}

\vspace{3mm}
\item[]\begin{tabular}{@{}ccc}
\br
  Model  &       $<Q_{KS}>$       &   $<D>$  \\
\mr
 RP01    &   $5.38\cdot 10^{-1}$  &   0.037  \\
 RP11    &   $6.83\cdot 10^{-3}$  &   0.097  \\
 RP12    &   $1.11\cdot 10^{-3}$  &   0.103  \\
 RP21    &   $1.32\cdot 10^{-7}$  &   0.149  \\
 RP22    &   $1.63\cdot 10^{-8}$  &   0.164  \\
 SUGRA11 &   $1.44\cdot 10^{-1}$  &   0.059  \\
 SUGRA12 &   $1.32\cdot 10^{-1}$  &   0.061  \\
 SUGRA21 &   $3.56\cdot 10^{-2}$  &   0.077  \\
 SUGRA22 &   $3.49\cdot 10^{-2}$  &   0.079  \\
 GPV     &   $5.27\cdot 10^{-5}$  &   0.124  \\
\br
\end{tabular}
\end{indented}
\end{table}

\end{document}